\documentclass[prl,twocolumn,showpacs,superscriptaddress]{revtex4}
\usepackage{graphicx,bm,amssymb,amsmath}

\newcommand{\Eq}[1]{Eq.~\eqref{#1}}
\newcommand{\eq}[1]{\eqref{#1}}
\newcommand{\Fig}[1]{Fig.~\ref{#1}}

\newcommand{\beq}{\begin{equation}}
\newcommand{\eeq}{\end{equation}}

\newcommand{\beqa}{\begin{eqnarray}}
\newcommand{\eeqa}{\end{eqnarray}}

\newcommand{\Beqa}{\begin{eqnarray*}}
\newcommand{\Eeqa}{\end{eqnarray*}}

\newcommand{\nn}{\nonumber}

\newcommand{\pdag}{{\phantom{\dagger}}}

\newcommand{\E}{\mathrm{e}}

\newcommand{\PRL}[3]{Phys. Rev. Lett.~\textbf{#1}, #2 (#3)}

\newcommand{\PRA}[3]{Phys. Rev. A~\textbf{#1}, #2 (#3)}
\newcommand{\PRD}[3]{Phys. Rev. D~\textbf{#1}, #2 (#3)}

\newcommand{\RMP}[3]{Rev. Mod. Phys.~\textbf{#1}, #2 (#3)}

\newcommand{\Nature}[3]{Nature~\textbf{#1}, #2 (#3)}

\newcommand{\JETP}[3]{Sov. Phys. JETP~\textbf{#1}, #2 (#3)}

\newcommand{\JMP}[3]{J. Math. Phys.~\textbf{#1}, #2 (#3)}

\newcommand{\etal}{\textit{et al.}} 
\newcommand{\G}{\mathcal{G}}

\begin{document}

\title{
Relaxation of a high-energy quasiparticle in a one-dimensional Bose gas
}

\author{Shina Tan}
\affiliation{
Department of Physics, 
Yale University, New Haven, CT 06520, USA}

\author{Michael Pustilnik}
\affiliation{
School of Physics, 
Georgia Institute of Technology, Atlanta, GA 30332, USA
}

\author{Leonid I. Glazman}
\affiliation{
Department of Physics, 
Yale University, New Haven, CT 06520, USA}

\begin{abstract}
We evaluate the relaxation rate of high-energy quasiparticles in a
weakly-interacting one-dimensional Bose gas. Unlike in higher dimensions, 
the rate is a nonmonotonic function of temperature, with a maximum at the 
crossover to the state of suppressed density fluctuations. At the maximum, 
the relaxation rate may significantly exceed its zero-temperature value.
We also find the dependence of the differential inelastic scattering rate on
the transferred energy. This rate yields information about temperature dependence of local pair correlations. 

\end{abstract}

\pacs{ 03.75.Kk,05.30.Jp} 
\maketitle 

Recent experiments with ultracold atomic gases~\cite{BDZ,Davidson_RMP} 
renewed the interest in fundamental properties of the elementary excitations in interacting Bose systems.

A three-dimensional (3D) Bose gas undergoes the Bose-Einstein
condensation (BEC) phase transition at a sufficiently low
temperature~\cite{Pitaevskii_Stringari}. The transition affects
dramatically the spectrum of elementary excitations (quasiparticles)
of the system. In the Bose-condensed phase, the quasiparticles obey
the Bogoliubov dispersion relation $\epsilon_q=sq\sqrt{1+(q/2ms)^2}$
which interpolates between a phonon-like linear spectrum at small
momenta (here $s$ is sound velocity and $m$ is each boson's mass) and a free-particle-like spectrum
at large momenta~\cite{Pitaevskii_Stringari}.

The BEC transition affects strongly the lifetime of low-energy 
quasiparticles. The relaxation rate $\Gamma_q$ of quasiparticles in 
the phonon part of the spectrum is very sensitive to both their momenta 
$q$~\cite{Beliaev} and temperature $T$~\cite{HM,Popov,PS}, 
$\Gamma_q \propto \max\{(\varepsilon_q)^5, \varepsilon_q T^4\}$. 
However, the relaxation rate of high-energy quasiparticles is dominated 
by collisions with large momentum transfer, does not depend on either $q$ 
or $T$, and thus is not sensitive to BEC transition~\cite{Beliaev}.
Some of these long-standing predictions have been recently verified experimentally, 
see~\cite{BDZ,Davidson_RMP} for a review.

Unlike its 3D counterpart, the one-dimensional (1D) interacting Bose gas 
turns at low temperatures to a quasicondensate in which the long-range 
order is destroyed by quantum fluctuations~\cite{Pitaevskii_Stringari,Popov}, 
and the BEC transition turns to a crossover. Yet, despite this difference, 
the spectrum of elementary excitations in 1D is still described very well by the Bogoliubov dispersion relation~\cite{Lieb}. 

However, the quasiparticle lifetime in 1D is very different from that in higher 
dimensions and is not as well understood. The reason is that, due to the constraints 
imposed by the energy and momentum conservation, two-particle collisions do 
not lead to a relaxation in 1D.
At the same time, realizations of 1D Bose systems with cold atoms confined 
in tight atomic waveguides~\cite{BDZ} are described rather well~\cite{Olshanii} 
by a model of bosons with zero-range repulsive interaction (the Lieb-Liniger model), 
which is integrable~\cite{Lieb,3particle}. In this model, the redistribution of the 
momenta between particles in a collision, and, therefore, relaxation, are 
absent~\cite{3particle}. Such apparent lack of relaxation was recently
demonstrated experimentally~\cite{Weiss_cradle} (see the discussion below).

The leading corrections to the Lieb-Liniger model have the form of a local
three-particle interaction term~\cite{Muryshev,MSS}, which breaks the
integrability and brings about the quasiparticle relaxation. In this Letter, 
we study relaxation of a particle with a large momentum. This problem was
considered recently in~\cite{MSS}, where the inelastic relaxation rate 
due to three-particle collisions was evaluated in the approximation that 
neglects two-body repulsion. The results of Ref.~\cite{MSS} suggest that, 
very much like in 3D, the relaxation rate of high-energy quasiparticles is independent
of momentum and temperature. However, in the present Letter, we demonstrate that, 
in a dramatic departure from the behavior in higher dimensions, the relaxation 
rate in 1D depends strongly on temperature even at large momenta.
It has a pronounced peak at the crossover to the quasicondensate state.

We evaluate the differential and the total relaxation rates. Both can be inferred from 
observations of colliding clouds of cold atoms~\cite{Weiss_cradle,Davidson}. 

To describe the relaxation in a weakly-interacting 1D Bose gas, we
consider the simplest Hamiltonian
\beq
H = H_0 + V,
\label{model}
\eeq
where
\beq
H_0 = \!\int\!dx\,\psi^\dagger(x)\!
\left(-\frac{1}{2m}\frac{d^{\,2}}{d x^2}\right)\psi(x)
+ \,\frac{\,c}{2}\!\int\!dx\! :\!\rho^2(x)\!:
\label{H_0}
\eeq
describes 1D bosons with a repulsive contact interaction (hereinafter we set $k_B = \hbar = 1$), 
and
\beq
V = - \frac{\alpha}{9 m}\!\int\!dx :\!\rho^3(x)\!:
\label{V}
\eeq
represents the leading integrability-breaking perturbation~\cite{Muryshev,MSS}. 
In Eqs.  \eq{H_0} and \eq{V}, $\rho(x) = \psi^\dagger(x)\psi(x)$ is the local density
operator and the colons denote the normal ordering. The strength of the interaction 
[represented by the second term in \Eq{H_0}] is characterized~\cite{Lieb} by the 
dimensionless parameter $\gamma=mc/n$, where $n$ is the 1D concentration.  A finite
three-particle scattering amplitude appears already in the first order
in $\alpha\ll 1$.

In this Letter, we study relaxation of a boson %added to a single-particle state 
with momentum $q$ (we assume that $q>0$) and kinetic energy $\xi_q=q^2\!/2m$, which is large compared to both temperature $T$ and a typical interaction energy per particle $\omega_s$, 
\beq
\xi_q \gg \max\{T,\omega_s\},
\quad
\omega_s = ms^2\!/2.
\label{xi_q}
\eeq
In the limit of a weak interaction $\gamma\ll 1$, which we consider from now on,
the sound velocity $s$ in \Eq{xi_q} is given by~\cite{Lieb}
$s = (n/m)\sqrt\gamma$. The condition \eq{xi_q} ensures that the particle is 
added to an almost empty single-particle state: 
$f_q = \langle\psi^\dagger_q\psi^\pdag_q\rangle\ll 1$ \cite{second_order,OD}. 

In the lowest (second) order in $\alpha$ the differential rate of inelastic
scattering is given by
\beq
\sigma_q(\omega) 
= \frac{\alpha^2}{2\pi m^2}\!\int^{\,q/3}_{-\infty}
\!dp\, 
\delta\bigl(\omega - \xi_q +\xi_{q-p}\bigr)\,{\cal G}(p,\omega)\,,
\label{sigma}
\eeq
where ${\cal G}(p,\omega)=\int\! dx dt\, e^{i\omega t-ipx} {\cal
 G}(x,t)$ is the Fourier transform of the correlation function 
\beq
{\cal G}(x,t) = \bigl\langle :\!\rho^2(x,t)\!:\,:\!\rho^2(0,0)\!:\bigr\rangle,
\label{G}
\eeq
which should be evaluated for the Lieb-Liniger model \Eq{H_0}. 
In writing \Eq{sigma} we took into account the kinematic constraint $p<q/3$
on the momentum transfer in the course of three-particle scattering.
The constraint translates into a restriction on the transferred energy:
$\sigma_q(\omega)$ vanishes for $\omega>5\xi_q/9$.
In terms of $\sigma_q(\omega)$, the total relaxation rate is given by 
\beq
\Gamma_q = \int\!d\omega\,\sigma_q(\omega).
\label{Gamma}
\eeq

The differential rate \eq{sigma} at large energy transfer $\omega$ 
is determined by the behavior of ${\cal G}(x,t)$ at $t\to 0$,
\beq
{\cal G}(x,t) = 2n^2g_2\Psi^2,
\quad
\Psi(x,t) = \left(\frac{m}{2\pi i t}\right)^{1/2}\!
e^{imx^2\!/2t}.
\label{small_t}
\eeq
Here $n^2g_2 = \langle:\!\!\rho^2(0,0)\!\!:\rangle$ is the probability of finding two bosons
at point $x=0$ at time $t=0$, and $\Psi(x,t)$ is the solution of the single-particle 
Schr\"odinger equation with the initial condition $\Psi(x,0) = \delta(x)$. 
Interactions do not affect the time evolution in \Eq{small_t} as long as 
$|t|\ll\min\{1/\omega_s,1/T\}$. Instead, the dependence on temperature and on the 
interaction strength enters \Eq{small_t} via the normalized local pair correlation 
$g_2$. For the Lieb-Liniger model this quantity can be evaluated exactly~\cite{KGDS}. 
For a weak interaction $g_2$ increases monotonically with $T$ from $g_2=1$
at $T\ll T_s$ to $g_2 = 2$ at $T\gg T_s$~\cite{KGDS}, where we introduced a characteristic temperature scale 
\beq
T_s = \sqrt{\omega_s T_0}=ns\,;
\label{Ts}
\eeq
here $T_0=2n^2\!/m$ is the quantum degeneracy temperature.
(Note that $\omega_s\ll T_s\ll T_0$ for a weak interaction.) 

Substitution of \Eq{small_t} into \Eq{sigma} yields the differential rate at 
large positive energy transfer \cite{large omega}
\beq
\label{sigma large omega}
\sigma_q(\omega) =
\frac{\alpha^2T_0 g_2}{2\pi \sqrt{\xi_q\omega}}
\bigg[
\frac{1+\sqrt{1-\omega/\xi_q}}{(1\!-\! \omega/\xi_q)(1+3\sqrt{1\!-\! \omega/\xi_q})}
\bigg]^{1/2}.
\eeq
\Eq{sigma large omega} is applicable at $\max\{\omega_s,T\}\ll\omega<5\xi_{q\!}/9$. 
Away from the upper end of this interval, at $\omega\ll\xi_q$, \Eq{sigma large omega} 
reduces to
\beq
\sigma_q(\omega) = \frac{\alpha^2T_0 g_2}{2\pi\sqrt{2\xi_q\omega}}\,.
\label{tail}
\eeq

To further analyze $\sigma_q(\omega)$ at $|\omega|\ll\xi_q$, we note that in 
this range of $\omega$ the momenta $p$ contributing to the integral in \Eq{sigma}
are small, $|p|\ll q$, and it simplifies to
\beq
\sigma_q(\omega) = \frac{\alpha^2}{2\pi mq}\,\G(0,\omega).
\label{sigma1}
\eeq
It follows from the properties of $\G(p,\omega)$ that the differential rate 
\eq{sigma1} satisfies the detailed balance condition
\beq
\sigma_q(-\omega)=e^{-\omega/T}\sigma_q(\omega).
\label{db}
\eeq
\Eq{db} implies that while $\sigma_q(\omega)\approx\sigma_q(-\omega)$ 
at small energy transfers $|\omega|\ll T$, the differential rate is exponentially small
at large negative $\omega$. 

To gain further understanding of the differential rate at small momentum 
transfer, we consider first the regime of relatively high
temperatures $T\gg T_s$, when the interaction in \Eq{H_0} can
be neglected (except for very tiny energy transfers, see below). The correlation function in \Eq{sigma1} is then
easily evaluated resulting in
\begin{align}
\label{sigma T>>Ts}
\sigma_q(\omega&)=\frac{\,\alpha^2}{2\pi^3m q}\!
\int\!\prod_{i=1}^4dk_{i\,} f_{k_1}f_{k_2}(f_{k_3\!\!}+1)(f_{k_4\!\!}+1)
\\
&\times\delta(k_1+k_2-k_3-k_4)\,
\delta(\xi_{k_1\!}+\xi_{k_2\!}-\xi_{k_3\!}-\xi_{k_4\!}+\omega),
\nn
\end{align}
where $f_k$ is the Bose distribution. (\Eq{sigma T>>Ts} can also be derived 
by using Fermi's Golden Rule.)

At $T_s\ll T\ll T_0$ the chemical potential is given by
\beq
\mu=-\mu_0,
\quad 
\mu_0 = T^2\!/T_0\ll T.
\label{mu_0}
\eeq
At $|\omega|\ll T$ the differential rate is dominated by processes 
in which both the initial and the final states of the 
two low-energy particles involved in a collision belong to the part of the spectrum 
with high occupation numbers: $f_{k_i}\approx f_{k_i} +1\approx T/(\xi_{k_i}+\mu_0)\gg1$.
Evaluation of \Eq{sigma T>>Ts} with this approximation results in
\beq
\sigma_q(\omega)
=\alpha^2(T_0/\xi_q)^{1/2}(T_0/T)^3 F\bigl(|\omega|/\mu_0\bigr),
\quad
|\omega|\ll T.
\label{sigma_free}
\eeq
The analytical expression for the function $F(z)$ is somewhat cumbersome.
It is a monotonic function normalized as $\int_0^\infty F(z)dz=1/8$, with a 
power-law behavior at $z\gg 1$, $F(z) = (2\sqrt{2}/\pi)z^{-5/2}$, and a logarithmic 
asymptote $F(z) = (5/16\pi^2) \ln (8e^{-7/5}/z)$ at $z\to 0$. 

The logarithmic divergence at $\omega\to 0$ in \Eq{sigma_free} comes
from $k_1\approx k_2\approx k_3\approx k_4$ in the integral over
momenta in \Eq{sigma T>>Ts}, and is an artifact of the free-boson
approximation. The probability of scattering two bosons with close
momenta $k_1\approx k_2$ in the initial state is suppressed at $|k_1-k_2|\ll mc$.  (There is a similar suppression for the final states $k_{3,4}$.) The logarithmic
divergence in $\sigma_q(\omega)$ is thus regularized at
$|\omega|\lesssim \omega_s^2/T_0$ for $T\gg T_s$.

At $T_s\ll T\ll T_0$ and $\omega\gg\mu_0$, the main contribution to the integral
in \Eq{sigma T>>Ts} comes from $|k_{1,2}|\lesssim\sqrt{m\mu_0}$ and 
$|k_{3,4}|\sim \sqrt{m\omega}\gg |k_{1,2}|$. Neglecting $k_{1,2}$ and $\xi_{k_{1,2}}$
in the arguments of the delta-functions in \Eq{sigma T>>Ts}, we find
\beq
\sigma_q(\omega)=\frac{\alpha^2T_0}{2\pi\sqrt{2\xi_q|\omega|}}
\frac{g_2}{\bigl(1-e^{-\omega/2T}\bigr)^{2}}\,,
\quad
|\omega|\gg\mu_0,
\label{sigma T}
\eeq
where $g_2=2$ as appropriate for $T\gg T_s$. 
\Eq{sigma T} extrapolates between Eqs.~\eq{tail} and \eq{sigma_free}. 

With lowering the temperature, Eqs. \eq{sigma T>>Ts}-\eq{sigma_free}
become inadequate when $\mu_0(T)$ is of the order of the interaction energy 
per particle $\omega_s$, i.e., at $T\sim T_s$.
At $T\ll T_s$, however, the Bogoliubov approximation for the local
density operator becomes applicable~\cite{Castin}.  In this
approximation, excitations of a 1D Bose liquid are essentially free
phonons (Bogoliubov quasiparticles), described by the Hamiltonian $H_B
= \sum_k\varepsilon_k b^\dagger_k b^\pdag_k$ with $\varepsilon_k =
\sqrt{\xi_k (\xi_k + 4\omega_s)}$. In terms of phonons, the density operator
has the form
$\rho(x)= n + \sum_{k\neq 0} (n\xi_k/L\varepsilon_k)^{1/2}\bigl(b_k^\pdag +b_{-k}^\dagger\bigr)e^{ikx}$,
where $L$ is the size of the system. Using this representation, evaluation of \Eq{sigma1} is straightforward and yields
\beqa
&&\sigma_q(\omega) = \frac{\alpha^2T_0}{64\pi\sqrt{2\omega_s\xi_q}}
\Big(\frac{\omega/\omega_s}{1-\E^{-\omega/2T}}\Big)^2
\label{sigma_Bogoliubov}
\\
&&\quad\times\bigl[1+(\omega/4\omega_s)^2\bigr]^{-1/2}\!
\Big\{1 +\bigl[1+(\omega/4\omega_s)^2\bigr]^{1/2}\Big\}^{-3/2}.
\nn
\eeqa
At $|\omega|\gg \omega_s$, \Eq{sigma_Bogoliubov} reduces to \Eq{sigma T}
with $g_2=1$ appropriate for $T\ll T_s$. 
In fact, \Eq{sigma T} is valid at any $T\ll T_0$, 
provided that the energy transfer falls within the range 
$\max\{\mu_0,\omega_s\}\ll |\omega|\ll\xi_q$. 
For positive $\omega$ in this range, the validity of \Eq{sigma T} is due to 
the fact that the interaction has negligible effect~\cite{second_order} 
on the final states of the colliding particles 
($\xi_{k_{3}}\approx\xi_{k_4}\approx\omega/2\gg\omega_s$).
The applicability of \Eq{sigma T} for negative $\omega$ in the above range then follows from \Eq{db}. 

%%%%%%%%%%%%%%%%%%%%%%%%%%%%%%%%
\begin{figure}
\includegraphics[width=0.7\columnwidth]{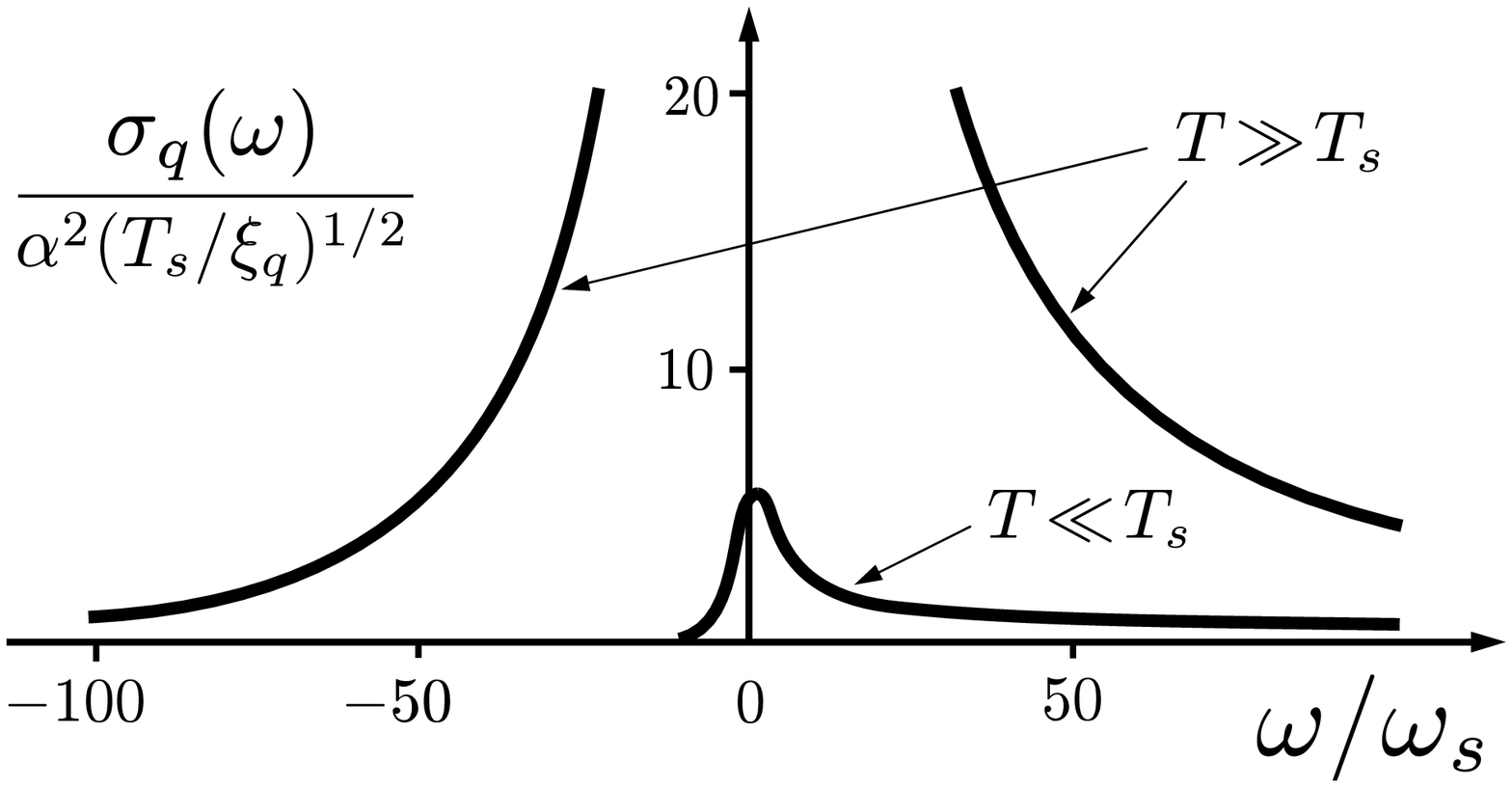}
\caption{
The differential inelastic scattering rate $\sigma_q(\omega)$ at different temperatures (only $|\omega|\ll\xi_q$ domain is shown).
The two plots correspond to Eqs. \eq{sigma T>>Ts} and \eq{sigma_Bogoliubov}, evaluated at $T =4T_s$  and $T = T_s/4$, respectively, with $T_s = T_0/16$. 
}
\label{fig1}
\end{figure}
%%%%%%%%%%%%%%%%%%%%%%%%%%%%%%%%

We show the typical plots of the differential relaxation 
rate $\sigma_q(\omega)$ at $T\gg T_s$ and $T\ll T_s$ in \Fig{fig1}.

We turn now to the evaluation of the total relaxation rate,
\Eq{Gamma}. There are two contributions to the integral over $\omega$
in \eq{Gamma}:
\beq
\Gamma_q = \Gamma_\infty + \widetilde\Gamma_q.
\label{Gamma1}
\eeq 
The first contribution, $\Gamma_\infty$, comes from the high-energy ``tail'' of 
$\sigma_q(\omega)$, see Eqs. \eq{sigma large omega} and \eq{tail}. This contribution 
is independent of $q$ and is given by~\cite{large omega}
\beq
\Gamma_\infty=\frac{\alpha^2T_0 g_2}{3\sqrt{3}}\,.
\label{background}
\eeq
Note that, unlike in higher dimensions, $\Gamma_\infty$ depends on temperature 
via $g_2(T)$, see the discussion above.

The second contribution in \Eq{Gamma1}, $\widetilde{\Gamma}_q\propto \xi_q^{-1/2}$, 
comes from the processes with a small energy transfer 
$|\omega|\lesssim\max\{T,\omega_s\}$. 
Using \Eq{sigma_free}, we find
\beq
\widetilde{\Gamma}_q = 
\frac{\alpha^2 T_0}{4}\, \left(\frac{T_s}{\xi_q}\right)^{\!1/2}\!\!
\left(\frac{T_0}{T_s}\right)^{\!3/2}\!\frac{\,T_s}{T}\,,
\quad
T_s\ll T\ll T_0.
\label{tau1}
\eeq
At lower temperatures we obtain, with the help of \Eq{sigma_Bogoliubov},
\beq
\widetilde{\Gamma}_q
= \frac{\alpha^2 T_0}{16}\left(\frac{T_s}{\xi_q}\right)^{\!1/2}\!\!
\left(\frac{T_0}{T_s}\right)^{\!3/2}\!\!\left(\frac{T}{T_s}\right)^{\!2}\!, 
\quad
\frac{T_s^2}{T_0}\ll T\ll T_s.
\label{tau3}
\eeq

\begin{figure}
\includegraphics[width=0.6\columnwidth]{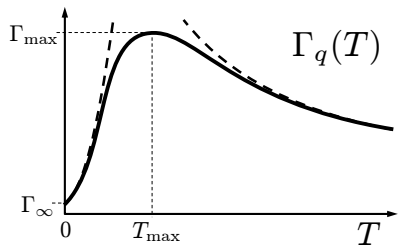}
\caption{
\label{fig2}
Sketch of the temperature dependence of the total inelastic relaxation rate.  
The dependence is nonmonotonic, with a maximum at $T_\text{max}\sim T_s$
The dashed lines indicate the high- and the low-temperature asymptotes 
Eqs.~\eq{Gamma1}-\eq{tau3}.
 }
\end{figure}

Comparison with \Eq{background} shows that for not too large energies,
$\xi_q\ll T_s(T_0/T_s)^3$, the small momentum transfer 
contribution $\widetilde{\Gamma}_q$ dominates the relaxation rate \eq{Gamma1} 
in a broad temperature interval
\beq\label{interval}
T_s\left(\frac{\xi_q}{T_s}\right)^{\!1/4}\!\!
\left(\frac{T_s}{T_0}\right)^{\!3/4}\!
\ll\,
T
\,\ll
T_s\!\left(\frac{T_s}{\xi_q}\right)^{\!1/2}\!\!
\left(\frac{T_0}{T_s}\right)^{\!3/2},
\eeq
which includes $T = T_s$. 
At some temperature $T_\text{max}\!\sim T_s$ within this interval, the relaxation 
rate reaches its peak value $\Gamma_\text{max} = \Gamma_q(T_\text{max})$, 
see \Fig{fig2}. 
By extrapolating the asymptotes \eq{tau1} and 
\eq{tau3} to the region $T\sim T_s$ and finding their intersection, we estimate 
$T_\text{max}\approx 1.6\, T_s$, and
\beq
\Gamma_\text{max} \approx 0.16\, \alpha^2 T_{0\,} (T_s/\xi_q)^{1/2}(T_0/T_s)^{3/2}.
\label{estimate}
\eeq
The actual values of $T_\text{max}$ and $\Gamma_\text{max}$ may differ from 
the above estimates only by numerical factors; finding these values would require a 
systematic description of the crossover regime $T\sim T_s$.

%%%%%%%%%%%%%%%%%%%%%%%%%%%
%%%%%%%%%%%%%%%%%%%%%%%%%%%%

We now discuss briefly the feasibility of observing relaxation by inelastic 
collisions in a system of cold atoms confined in a cylindrical trap. In this 
case the effective Hamiltonian \eq{model}-\eq{V} can be derived explicitly, 
by projection onto the lowest subband of transverse quantization.  For a
model in which the interaction in 3D is described by a pseudopotential
$V_{3D}(\bm r) = 4\pi (a/m)\delta(\bm r)$, where $a$ is the $s$-wave
scattering length~\cite{Pitaevskii_Stringari}, and with the amplitude of radial 
zero-point motion $a_r = (m\omega_r)^{-1/2}\gg a$ (here $\omega_r$ is the 
trap frequency), one finds~\cite{Olshanii,MSS,MSS4}
\beq
\gamma=2a/n a_r^2, 
\quad
\alpha=18\ln(4/3)(a/a_r)^2.
\label{alphagamma}
\eeq 

The main limitation arises due to 3-body recombination
processes~\cite{recombination}, absent in our model. The corresponding
rate is $\Gamma_R =\beta n^{2\!}
g_3/a_r^4$~\cite{recombination}, where $g_3 = \langle
:\!\rho^3\!\!:\rangle/n^3$.
Using Eqs. \eq{background} and \eq{alphagamma},
we find~\cite{large omega}
\beq
\Gamma_\infty/\Gamma_R =\eta g_2/g_3,
\quad
\eta =10.3a^4/(m\beta)\,.
\label{infty}
\eeq
For $^{87}$Rb
($a = 5.3\,\text{nm}$,
$\beta =3\times 10^{-31}\,\text{cm}^6\!/\text{s}$ \cite{Burt}),
we have $\eta\approx20$.  For a weak to a moderately strong interaction, 
$\gamma\lesssim 1$, the ratio $g_2/g_3$ in \Eq{infty} is of the order of 1 at all $T$, 
and $\Gamma_\infty/\Gamma_R\sim 10$. 

In experiments with periodically colliding clouds 
of cold gases~\cite{Weiss_cradle}, the 3-body recombination occurs 
all the time, while the scattering between the clouds takes place only during 
the collision itself (about one tenth of a period in~\cite{Weiss_cradle}). Therefore, 
the probability that during a period a particle participates in an inelastic collision 
event with a large energy transfer, and the probability that it participates in a 3-body
recombination process are of the same order. Accordingly,  inelastic 
scattering with a large energy transfer is difficult to detect unambiguously.

Relaxation by the inelastic scattering with a small energy transfer is effective 
when the interaction is weak, $\gamma\ll 1$ 
(indeed, the interval $T_s^2\!/T_0\ll T\ll T_0$, see 
Eqs. \eq{tau1} and \eq{tau3}, disappears for large $\gamma$). 
For the peak value of $\Gamma_q$ [see \Eq{estimate}], we find
\beq
\Gamma_\text{max}/\Gamma_R \sim 2.3\,\eta (T_s/\xi_q)^{1/2}\gamma^{-3/4},
\label{max}
\eeq
which for a fixed ratio $\xi_q/T_s$ diverges in the limit $\gamma\to 0$.

The maximum of $\Gamma_q$ is reached at $T\sim T_s$. 
The condition of the observability of the inelastic relaxation,
$\Gamma_\text{max}\gg \Gamma_R$, and the condition for the
high-energy quasiparticle to be outside the quasicondensate yet well within the lowest 
subband of transverse quantization, $T_s\ll\xi_q\ll\omega_r$, can be satisfied simultaneously. For example, for $^{87}$Rb the trap frequency $\omega_r/2\pi = 15\,\text{kHz}$ 
and concentration $n = 7\,\mu\text{m}^{-1}$  
correspond to $\gamma = 0.2$ and $T_s = 120\,\text{nK}$. 
For $\xi_q/T_s = \omega_r/\xi_q = 2.4$, \Eq{max} then yields 
$\Gamma_\text{max}/\Gamma_R\sim100$. The parameters above are realistic 
with today's experimental technology~\cite{Weiss_cradle,Schmiedmayer}.
 
In conclusion, we evaluated the quasiparticle relaxation rate in a weakly-interacting 1D Bose liquid.  Unlike in 3D, the rate is strongly momentum and temperature
dependent, with a maximum at $T\sim ns$, where $s$ is the
sound velocity at $T=0$. Our predictions can be verified in experiments with colliding clouds of cold atoms.

\begin{acknowledgments}
We acknowledge discussions with V. Cheianov, D. Gangardt, R. Hulet,
A. Kamenev, A. Kuzmich, G. Shlyapnikov, and D. Weiss, hospitality of Kavli
Institute for Theoretical Physics at UCSB and Aspen Center for
Physics, and support by the DOE (Grant No. DE-FG02-06ER46311) and by the NSF 
(Grant No. DMR-0906498).
\end{acknowledgments}


\begin{thebibliography}{99}
\bibitem{second_order}
For a weak interaction, the momentum distribution at 
$\xi_k\gg \max\bigl\{\omega_s,T\bigr\}$ can be obtained perurbatively:
$f_k = dE_2\!/d\xi_k\sim (nc/\xi_k)^2\sim (\omega_s/\xi_k)^2$, where $E_2$ 
is the second-order correction to the ground state energy.
\bibitem{large omega}
Eqs. \eq{sigma large omega}, \eq{background}, \eq{infty} are valid 
for all $\gamma$, not just $\gamma\ll 1$.
\bibitem{MSS4}
Our result for $\alpha$ is 4 times smaller than that of Ref.~\cite{MSS}.
\end{thebibliography}
\end{document}